\newcommand{\be}{\begin{equation}}
\newcommand{\ee}{\end{equation}}
\newcommand{\bea}{\begin{eqnarray}}
\newcommand{\eea}{\end{eqnarray}}

\documentclass[twocolumn,showpacs,preprintnumbers,amsmath,amssymb]{revtex4}

\usepackage{graphicx}
\usepackage{dcolumn}
\usepackage{bm}

\begin{document}

\preprint{APS/123-QED}

\title{Exact static solutions to a translationally invariant discrete
$\phi^4$ model}

\author{Sergey V. Dmitriev$^1$, Panayotis G. Kevrekidis$^2$, Avinash Khare$^3$, and
Avadh Saxena$^4$}
\affiliation{$^1$ Institute of Industrial Science, the University
of Tokyo, Komaba, Meguro-ku, Tokyo 153-8505, Japan \\
$^2$ Department of Mathematics and Statistics,
University of Massachusetts, Amherst, MA 01003-4515, USA \\
$^3$Institute of Physics, Bhubaneswar, Orissa 751005,
India \\
$^4$ Center for Nonlinear Studies and Theoretical Division, Los
Alamos National Laboratory, Los Alamos, New Mexico 87545, USA}

\date{\today}

\begin{abstract}
For a discrete, translationally-invariant $\phi^4$ model
introduced by Barashenkov {\it et al.} [Phys. Rev. E {\bf 72},
35602R (2005)], we provide the momentum conservation law and
demonstrate how the first integral of the static version of the
discrete model can be constructed from a Jacobi elliptic function
(JEF) solution. The first integral can be written in the form of a
nonlinear map from which {\it any} static solution supported by
the model can be constructed. A set of JEF solutions, including
the staggered ones, is derived. We also report on the stability
analysis for the static bounded solutions and exemplify the
dynamical behavior of the unstable solutions. This work provides a
road-map, through this illustrative example, on how to fully analyze
translationally-invariant models in terms of their static problem,
its first integral, their full-set of static solutions and
associated conservation laws.
\end{abstract}


\maketitle

\section{Introduction and Setup} \label{Introduction}

In recent years, the discrete nonlinear models have played a very
important role in several physics applications \cite{Toda,braun}.
The question of mobility of solitonic excitations in discrete media
is a key issue in many physical contexts; for example the mobility of
dislocations, a kind of topological solitons, is of importance in the
physics of plastic deformation of metals and other crystaline bodies
\cite{Nabarro}. Similar questions arise in optics
for light pulses moving in optical
waveguides or in photorefractive crystal lattices (see e.g.,
\cite{melvin} for a relevant recent discussion) and in atomic
physics for Bose-Einstein condensates moving through optical
lattice potentials (see e.g., \cite{oberthaler} for a recent review).
These issues may prove critical in aspects related
to the guidance and manipulation of coherent, nonlinear wavepackets
in solid-state, atomic and optical physics applications.

In the above settings, the translationally invariant
(TI) discrete models \cite{PhysicaD}
have received considerable attention since they
admit static solutions that can be placed anywhere with respect
to the lattice.  For the Hamiltonian TI lattices
\cite{SpeightKleinGordon,CKMS_PRE2005}, this can be interpreted as
the absence of the Peierls-Nabarro potential \cite{Nabarro}. For the
non-Hamiltonian lattices, the height of the Peierls-Nabarro barrier
is path-dependent but there exists a continuous path along which
the work required for a quasi-static shift of the solution along the
lattice is zero \cite{DKYF_PRE2006}. In general, one can state that
coherent structures in the TI models are not trapped by the
lattice and they can be accelerated by even a weak external field.
This particular property makes the TI discrete models potentially
interesting for physical applications and one such physically
meaningful model has been recently reported \cite{Coulomb}.

It has been demonstrated that some of the TI lattices support
traveling solutions, but only for selected velocities
\cite{oxt1,DNLSE3}; this feature distinguishes them from the
completely integrable lattices \cite{Toda,AL,Orfanidis} where the
propagation velocity can change continuously. Translationally
invariant discrete models can also be defined as models whose
static version is exactly solvable; thus, the TI discrete models
are directly connected to the theory of integrable maps
\cite{Quispel}. Starting from the pioneering works of
\cite{SpeightKleinGordon,PhysicaD}, the TI discrete models have
been receiving attention from several groups. General classes of
such models have been constructed and investigated for the
Klein-Gordon
\cite{Submitted,CKMS_PRE2005,JPhysA,DKYF_PRE2006,Coulomb,oxt1,Barashenkov}
and for the nonlinear Schr{\"o}dinger equations (NLSE)
\cite{DNLSE1,DNLSE2,KhareJPA,KharePRE,DNLSE3,DEPelinovsky}.

For a wide class of Klein-Gordon TI models it is known that they
conserve momentum defined in a standard form \cite{PhysicaD} but
are non-Hamiltonian. That is to say, there are TI models that
conserve the standard  momentum but not the standard energy.
Similarly, there are models that conserve energy but do not
conserve the momentum; see \cite{Submitted}, for a detailed
discussion. However, this raises the question: is there always
some form of a conservation law (either as a standard momentum or
a standard energy, or as a generalized version thereof) present in
the TI models? In this paper, we consider an example of a model
that conserves neither the standard energy, nor the standard
momentum. We show in this case that a conservation of a
non-standard momentum is present. The existence of a conservation
law even in such a non-standard case, leads us to believe that
there should always be at least one conservation law in the TI
models.

It is also known that some of the discrete TI models, both of the
Klein-Gordon and of the NLSE type, support exact static/stationary
and sometimes even moving (with a discrete set of velocities)
solutions in the form of the Jacobi elliptic functions (JEF)
\cite{CKMS_PRE2005,DKYF_PRE2006,KhareJPA,KharePRE,DNLSE3} but
other TI lattices do not support JEF solutions, e.g., the lattice
reported in \cite{SpeightKleinGordon}.

On the other hand, for any TI lattice, the static problem can be
reduced to a first-order difference equation which has been called
the discretized first integral (DFI) \cite{JPhysA,DKYF_PRE2006}
because, in the continuum limit, the DFI reduces to the first
integral of the static field equation. The DFI is a nonlinear map from
which, for an admissible initial value, one can construct a static
solution iteratively.
The DFI approach can be used to construct the TI models
\cite{JPhysA,DKYF_PRE2006}, but here we address the inverse
problem of finding the DFI for a given TI model. We demonstrate how
the DFI can be found from a known JEF solution to a TI model.

More specifically, in the present study we consider the following
discrete model
first suggested by Barashenkov {\it et al.} \cite{Barashenkov},
\begin{eqnarray} \label{Barashenkov}
   \ddot{\phi}_n=\frac{1}{{h^2 }}\left( {\phi _{n - 1}  - 2\phi _n  + \phi _{n + 1}
   } \right) + \lambda \phi _n  
   - \lambda \phi _{n - 1} \phi _n \phi _{n + 1},
\end{eqnarray}
where $\lambda= \pm 1$ and overdot means derivative with respect
to time. In the continuum limit $(h \rightarrow 0)$ the model
reduces to the $\phi^4$ field theory model
\begin{equation} \label{KleinGordon}
   \phi _{tt}  = \phi _{xx} +\lambda\left(\phi - \phi^3\right).
\end{equation}
The static version of Eq. (\ref{KleinGordon}) (with omitted inertia
term $\phi _{tt}$) possesses the first integral
\begin{eqnarray} \label{FirstInt}
   \phi_x^2 - \frac{\lambda}{2}(1-\phi^2)^2 + K = 0 \,,
\end{eqnarray}
where $K$ is the integration constant. An exact static kink
solution to the TI lattice of Eq. (\ref{Barashenkov}) and a
two-point map that generates the kink solution were derived by
Barashenkov {\it et al.} \cite{Barashenkov}.

There are numerous reasons for the choice of Eq.
(\ref{Barashenkov}) as our illustrative example for this study, intending
to highlight the range and the power of the methods available to
tackle TI models. Firstly, the model does not possess an obvious
conservation law. Secondly, it does not have an obvious discrete
first integral. Thirdly, its general solutions are not available
and neither their parametric space, nor their stability have
been explored. In the present
work, we develop a number of relevant results for the lattice of
Eq. (\ref{Barashenkov}). In particular, in Sec. \ref{Analytics},
we give (i) the corresponding conservation law; (ii) the two-point
map (or DFI) from which {\em any} static solution of Eq.
(\ref{Barashenkov}) can be constructed; (iii) a set of static JEF
solutions. We believe that the features (i) and (ii) should be
present in any TI model, and use this non-standard model to
showcase how to develop them. Then, in Sec. \ref{Numerics}, we
present the results of numerical studies including the stability
analysis of the static solutions and, for $h^2<2$, we show several
examples of the dynamics of unstable solutions. Section
\ref{Conclusions} concludes the paper. It is worth pointing out
that recently two of us \cite{DKYF_PRE2006} have similarly studied
a model (henceforth called the --standard-- momentum conserving or
MC model) \be\label{mc} \ddot{\phi}_n=\frac{1}{{h^2 }}\left( {\phi
_{n - 1} - 2\phi _n + \phi _{n + 1}
   } \right) + \lambda \phi _n  
   - \frac{\lambda}{2}\phi_n^2[ \phi _{n - 1}+\phi _{n + 1}]\,.
\ee We shall compare and contrast the various results obtained
from model of Eq. (\ref{Barashenkov}) with those of the MC model to get a
better insight into the various issues involved.

\section{Analytical results} \label{Analytics}

\subsection{Momentum conservation law} \label{Momentum}

By seeking expressions that would symmetrize the form of the nonlinearity,
we have found that the discrete TI model of Eq. (\ref{Barashenkov})
conserves the momentum defined as
\begin{eqnarray} \label{mom2}
   P=\sum_n \dot{\phi}_n (\phi_{n+2}-\phi_{n-2}),
\end{eqnarray}
which is different from the standard definition of momentum used,
e.g., in \cite{PhysicaD,Submitted}. The latter momentum, namely:
\begin{eqnarray} \label{mom3}
   P=\sum_n \dot{\phi}_n (\phi_{n+1}-\phi_{n-1}),
\end{eqnarray}
is conserved e.g. by the
 MC model
of Eq. (\ref{mc}). The proof in both cases is through straightforward
time-differentiation and application of a  telescopic summation.

\subsection{Reduction to the first-order difference equation} \label{TwoPoint}

It is possible to construct the first integral of the static
version of Eq. (\ref{Barashenkov}), which can be presented in the
form of the nonlinear map,
\begin{eqnarray} \label{Two_point_map}
   \phi_{n \pm 1}  &=& (2 - \Lambda)\frac{Z\phi_n
   \pm \sqrt {f(\phi_n)} }{2 - \Lambda  - \Lambda \phi_n^2 }, \nonumber \\
   f(\phi _n) &=& \frac{\Lambda }{2 - \Lambda}
   \left( {\tilde K - X\phi _n^2  + \phi _n^4 } \right),
\end{eqnarray}
with
\begin{equation} \label{Lambda}
   \Lambda=\lambda h^2,
\end{equation}
\begin{equation} \label{KFarbtilde}
   \tilde{K} = 1 - \frac{2K}{\lambda},
\end{equation}
\begin{eqnarray} \label{Z}
   Z &=& \frac{(2 - \Lambda)^2 - \tilde K\Lambda ^2}
   {2(2 - \Lambda)},
\end{eqnarray}
\begin{eqnarray} \label{X}
   X &=& \frac{\tilde K\Lambda ^2  + (2 - \Lambda)^2 (1 - Z^2)}
   {\Lambda (2 - \Lambda)}.
\end{eqnarray}

Apart from the model parameters, $h^2$ and $\lambda$, the map
defined by Eqs. (\ref{Two_point_map}) to (\ref{X}) contains the
integration constant $K$. The role of the second integration
constant, for the second-order difference static equation
(\ref{Barashenkov}), is effectively played by the initial value of
the map, $\phi_0$. In Appendix I we show how the two-point map can
be constructed from the known JEF solutions. Due to the symmetry
of Eq. (\ref{Barashenkov}), in the left-hand side of Eq.
(\ref{Two_point_map}), one can substitute $\phi_{n \pm 1}$ with
$\phi_{n \mp 1}$.

For any pair of admissible values, $K$ and $\phi_0$, the map of
Eq. (\ref{Two_point_map}) generates a static solution to Eq.
(\ref{Barashenkov}). Since the same map is used for calculation of
the forth and back points of the map, iterations starting from an
admissible value $\phi_0$ cannot give an inadmissible one and
thus, the static solution will necessarily be constructed for the
entire chain. From the two different roots of Eq.
(\ref{Two_point_map}) for $\phi_{n+1}$ (or for $\phi_{n-1}$), one
should take the one different from $\phi_{n-1}$ (or from
$\phi_{n+1}$). This rule gives the sufficient condition for moving
along the same solution branch and it guarantees that, for the
chosen $\phi_{n+1}$ (or for $\phi_{n-1}$), the three point problem
Eq. (\ref{Barashenkov}) will be satisfied for $\phi_n$.

In the continuum limit, the first-order difference equation
(\ref{Two_point_map}) reduces to Eq. (\ref{FirstInt}). In other
words, the two-point map, as given by Eq. (\ref{Two_point_map}),
is a DFI of the
static $\phi^4$ field. This observation is in line with our
earlier work \cite{JPhysA,DKYF_PRE2006} where the relation
between TI discrete models and DFI of the static field equation
was established.

It is interesting to note that for the MC model (\ref{mc}), the
nonlinear map has the form \cite{DKYF_PRE2006}
\begin{eqnarray} \label{mcmap}
   \phi_n =\frac{(2-\Lambda)\phi_{n+1}\pm\sqrt{D}}{(2-\Lambda
   \phi_{n+1}^2)}\,,
\end{eqnarray}
where
\begin{eqnarray} \label{mcD}
   D=2\Lambda(1-\phi_{n+1}^2)^2+2\tilde{C}(\Lambda\phi_{n+1}^2-2)\,,
\end{eqnarray}
and $\tilde{C}=Ch^2$ with $C$ being the integration constant.

It is also worth pointing out that while the map for the MC case
given by Eqs. (\ref{mcmap}) and (\ref{mcD}), is quadratic in both
$\phi_n$ and $\phi_{n+1}$, the map in the present case as given by
Eq. (\ref{Two_point_map}) is, generally speaking, quartic in
$\phi_n$ and quadratic in $\phi_{n+1}$, the only exception being
when $\Lambda=2$ when the map is quadratic in both $\phi_n$ and
$\phi_{n+1}$ (this is discussed in detail in Sec. \ref{Lambda2}).
Thus, while the map in the MC case belongs to the family of
twelve-parameter two-point map presented by Quispel {\it et al.}
\cite{Quispel}, the map in our case as given by Eq.
(\ref{Two_point_map}) does not, in general, belong to the relevant
family of maps. To be more precise, while the static version of
Eq.  (\ref{Barashenkov}) is of the Quispel {\it et al.}
\cite{Quispel} form as given by
\begin{eqnarray}
   \phi_{n+1}=\frac{f_1(\phi_n)-f_2(\phi_n)\phi_{n-1}}
   {f_2(\phi_n)-f_3(\phi_n)\phi_{n-1}}\,,
\end{eqnarray}
with
\begin{eqnarray}
   f_1(\phi_n)=(2-\Lambda)\phi_n\,,~~f_2(\phi_n)=1\,,
   ~~f_3(\phi_n^2)=\Lambda \phi_n\,,
\end{eqnarray}
one is unable to write the corresponding invariant map in terms of
the twelve parameters presented in \cite{Quispel}.

It should however be noted that, even then, as shown below,
all static solutions to Eq. (\ref{Barashenkov}) can
be obtained iteratively from Eq. (\ref{Two_point_map}) with an
admissible $K$, starting from any admissible $\phi_0$.

\subsection{Admissible values of integration constants \\
and solutions from the nonlinear map} \label{LinearStability}

Throughout this paper, without any loss of generality, we will
always take $\lambda=\pm 1$ and vary $h$ so as to vary $\Lambda$.

Inadmissible values of integration constants, $K$ and $\phi_0$,
are those for which, in Eq. (\ref{Two_point_map}), $f(\phi_0)<0$
and $2-\Lambda-\Lambda\phi_0^2=0$. For $\lambda=1$, we need to
consider separately the regime of $h^2<2$ (i.e. $0 < \Lambda <2$)
and the regime of very high discreteness, $h^2>2$ (i.e. $\Lambda
>2$). The third case is the case of $\Lambda <0$ where, however,
the topology does not change as one goes from small to large
$h^2$. For $\Lambda=2$ the two-point map Eq. (\ref{Two_point_map})
cannot be used in this form. This case is studied separately in
Sec. \ref{Lambda2}. It may be pointed out that similar studies
have already been performed for the MC model in
\cite{DKYF_PRE2006}.

\subsubsection{\it Case of $\,0 < \Lambda <2$}

\begin{figure}
\includegraphics{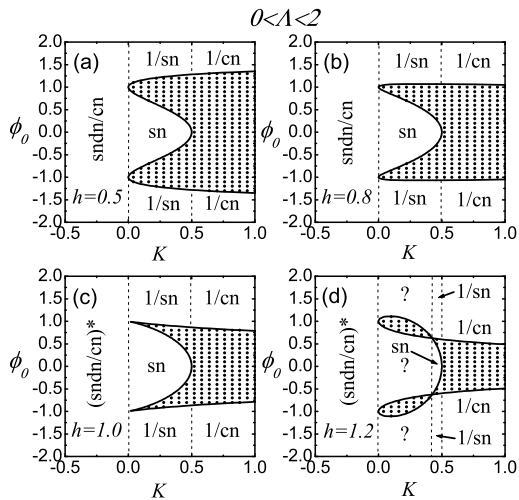}
\caption{Admissible and inadmissible (shaded) regions for the
integration constants $K$ and $\phi_{0}$ for $\lambda=1$ and
different values of $h<\sqrt{2}$, indicated in each panel. The
topology of the admissible regions is different for $h<1$ and
$1<h<\sqrt{2}$. We identify the portions of the admissible region
occupied by different JEF solutions and mark the staggered
solutions with $"*"$ (see Sec. \ref{Jacobi}). The only bounded
solution in this case is the sn solution which exists for
$0<K<1/2$ when $h<1$ and in a narrower range of $K$ otherwise. For
example, in panel (d), the sn solution exists for $0.425<K<1/2$.
The kink solution, corresponding to $K=0$ and $-1<\phi_0<1$,
exists only for $h<1$. The solution $({\rm sn/cn})^*$ cannot be seen
in (a)-(d) because it occupies the range of $-\infty < K <
\bar{K}$ with negative $\bar{K}$ being outside the frames of the
panels. In (a) and (b), the solution $({\rm sndn/cn})^*$ takes a
portion of $K$ in between $({\rm sn/cn})^*$ and ${\rm sndn/cn}$
solutions, but it falls out of the frame of these panels. For the
area marked with the question mark in (d) we could not identify a
JEF solution. Several static solutions generated by the map of Eq.
(\ref{Two_point_map}) for $h=0.8$, panel (b), are presented in
Fig. \ref{Figure2}.} \label{Figure1}
\end{figure}

The inadmissible regions of the plane $(K,\phi_0)$ are shown by
shaded areas in Fig. \ref{Figure1} for different values of the
discreteness parameter, indicated in each panel. The topology of
the admissible regions is different for $h<1$ and $1<h<\sqrt{2}$.
We identify the portions of the admissible region occupied by
different JEF solutions and mark the staggered solutions with
$"*"$ (see Sec. \ref{Jacobi}).

Admissible regions for $0 < \Lambda <1$ [see in (a) and (b) of
Fig. \ref{Figure1}] are completely filled by the six JEF
solutions: $({\rm sn/cn})^*$, $({\rm sndn/cn})^*$, ${\rm
sndn/cn}$, ${\rm sn}$, ${\rm 1/sn}$, and ${\rm 1/cn}$ (see Sec.
\ref{Jacobi}).

In the range of $1 < \Lambda <2$ [see in Fig. \ref{Figure1}(d)]
the solution ${\rm sndn/cn}$ disappears, while the remaining five
solutions persist and, there appears a portion of the admissible
region where we could not identify a JEF solution [marked with the
question mark in (d)]. However, this solution can be easily
obtained from the map Eq. (\ref{Two_point_map}) and one finds that
the solution is unbounded.

The solution $({\rm sn/cn})^*$ cannot be seen in (a)-(d) of Fig.
\ref{Figure1} because it occupies the range of $-\infty < K <
K^\ast$ with negative $K^\ast$ falling outside the frames of the
panels. In (a) and (b) of Fig. \ref{Figure1}, the solution $({\rm
sndn/cn})^*$ takes a portion of $K$ in between $({\rm sn/cn})^*$
and ${\rm sndn/cn}$ solutions, but it falls out of the frames of
these panels.

The only bounded solution in this case is the $\rm{sn}$ solution,
which exists for $0<K<1/2$ when $h<1$ and in a narrower range of
$K$ otherwise. For example, in panel Fig. \ref{Figure1}(d), the
$\rm{sn}$ solution exists for $0.425<K<1/2$. The kink solution,
Eq. (\ref{kink}), corresponding to $K=0$ and $-1<\phi_0<1$, exists
only for $h<1$ (i.e. only for $0 < \Lambda <1$).

\begin{figure}
\includegraphics{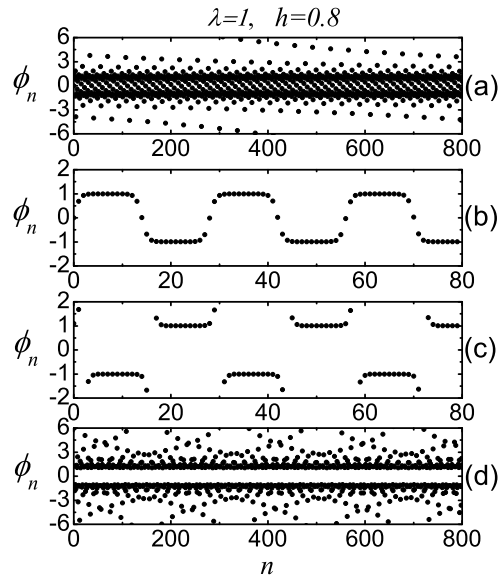}
\caption{Static solutions generated by the map Eq.
(\ref{Two_point_map}) for $\lambda=1$, $h=0.8$ [see panel (b) of
Fig. \ref{Figure1}] for different values of the integration
constants, $K$ and $\phi_0$. We have: (a) ${\rm sndn/cn}$ solution
($K=-0.25$, $\phi_0=0$); (b) ${\rm sn}$ solution close to the
hyperbolic limit ($K=10^{-8}$, $\phi_0=0$); (c) ${\rm 1/sn}$
solution close to the hyperbolic limit ($K=10^{-8}$,
$\phi_0=1.1$); and (d) ${\rm 1/cn}$ solution ($K=0.75$,
$\phi_0=1.25$). The only bounded solution is the ${\rm sn}$
solution shown in (b). Unbounded solutions have points with very
large amplitudes falling outside the frames of the figures.}
\label{Figure2}
\end{figure}

Several static solutions generated by the map
(\ref{Two_point_map}) for $h=0.8$ [see panel (b) of Fig.
\ref{Figure1}] are presented in Fig. \ref{Figure2}. In Fig.
\ref{Figure2}(a) we have the ${\rm sndn/cn}$ solution generated
with $K=-0.25$, $\phi_0=0$; in Fig. \ref{Figure2}(b), the ${\rm
sn}$ solution close to the hyperbolic limit generated with
$K=10^{-8}$, $\phi_0=0$; in Fig. \ref{Figure2}(c), the ${\rm
1/sn}$ solution close to the hyperbolic limit obtained for
$K=10^{-8}$, $\phi_0=1.1$; and in Fig. \ref{Figure2}(d), the ${\rm
1/cn}$ solution constructed for $K=0.75$, $\phi_0=1.25$. The only
bounded solution is the ${\rm sn}$ solution shown in Fig.
\ref{Figure2}(b), when it assumes the form of alternating kinks
and anti-kinks. For unbounded solutions, the amplitudes of some
points are very large and they fall outside the frames of the
figures.

\subsubsection{\it Case of $\,\Lambda>2$}

\begin{figure}
\includegraphics{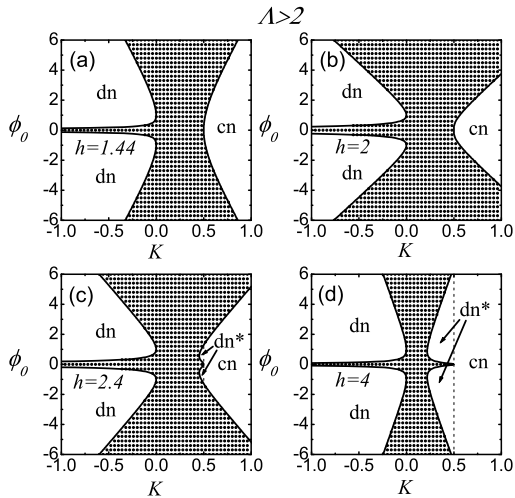}
\caption{Same as in Fig. \ref{Figure1} but for a regime of high
discreteness, $h^2>2$. A qualitative change of the border between
admissible and inadmissible regions takes place at $h=2$ when the
curvature of the border at the point $(K,\phi_0)=(1/2,0)$ changes
sign. All static solutions are bounded. Admissible regions are
completely filled by the three JEF solutions, ${\rm cn}$, ${\rm
dn}$, and staggered ${\rm dn}$, denoted as ${\rm dn}^*$ (see Sec.
\ref{Jacobi}). Several static solutions generated by the map of Eq.
(\ref{Two_point_map}) for $h=4$, panel (d), are presented in Fig.
\ref{FigureX}.} \label{Figure3}
\end{figure}

The inadmissible regions are shown by the shaded areas in Fig.
\ref{Figure3} for different values of $h$, indicated in each
panel. These are already quite different in structure from the
inadmissible regions of Fig. \ref{Figure1}; notice in particular
the existence of such regions for positive $K$, near $\phi_0=0$,
while in Fig. \ref{Figure1}, admissible $\phi_0$ exist for any
$K$. Notice also a qualitative change of the border between
admissible and inadmissible regions that takes place at $h=2$ when
the curvature of the border at the point $(K,\phi_0)=(1/2,0)$
changes sign. One can see that, in this case, all static solutions
are bounded. Admissible regions are completely filled by the three
JEF solutions, ${\rm cn}$, ${\rm dn}$, and staggered ${\rm dn}$,
denoted as ${\rm dn}^*$ (see Sec. \ref{Jacobi}).

\begin{figure}
\includegraphics{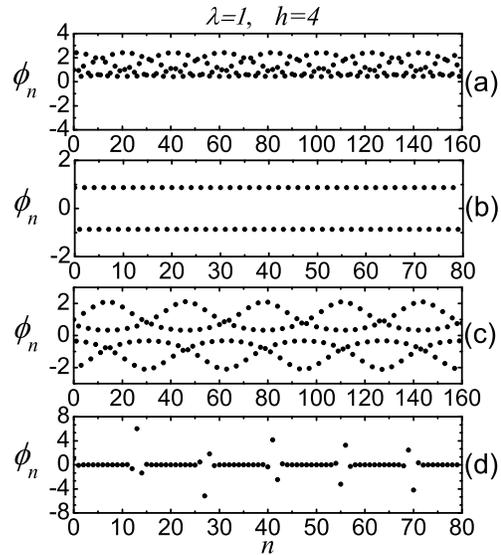}
\caption{Static solutions generated by the map of Eq.
(\ref{Two_point_map}) for $\lambda=1$, $h=4$ [see panel (d) of
Fig. \ref{Figure3}] for different values of the integration
constants, $K$ and $\phi_0$. All solutions are bounded. We have:
(a) ${\rm dn}$ solution ($K=-0.05$, $\phi_0=1$); (b) zigzag
solution, or the ${\rm dn}^*$ solution in the limit of $m=0$
($K=7/32$, $\phi_0=\sqrt{3}/2$); (c) ${\rm dn}^*$ solution
($K=0.26$, $\phi_0=1$); and (d) the ${\rm cn}$ solution with $K(m)
< \beta h <2K(m)$ close to the hyperbolic limit, where it obtains
the form of the chain of staggered pulses, Eq.
(\ref{pulse_stagg}),  ($K=1/2+10^{-12}$, $\phi_0=1$).}
\label{FigureX}
\end{figure}

Examples of static solutions generated by the map of Eq.
(\ref{Two_point_map}) for $\lambda=1$, $h=4$ [corresponds to panel
(d) of Fig. \ref{Figure3}] for different values of the integration
constants, $K$ and $\phi_0$, are plotted in Fig. \ref{FigureX}.
The solutions are: (a) ${\rm dn}$ solution ($K=-0.05$,
$\phi_0=1$); (b) zigzag solution, which is actually the ${\rm
dn}^*$ solution in the limit of $m=0$ ($K=7/32$,
$\phi_0=\sqrt{3}/2$); (c) ${\rm dn}^*$ solution ($K=0.26$,
$\phi_0=1$); and (d) ${\rm cn}$ solution with $K(m) < \beta h
<2K(m)$ close to the hyperbolic limit, where it assumes the form
of the chain of staggered pulses, Eq. (\ref{pulse_stagg}),
($K=1/2+10^{-12}$, $\phi_0=1$).

\subsubsection{\it Case of $\Lambda < 0$}

\begin{figure}
\includegraphics{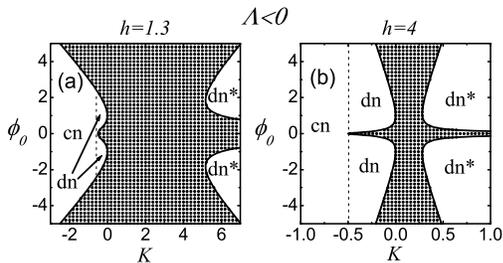}
\caption{Same as in Fig. \ref{Figure1} and Fig. \ref{Figure3} but
for $\lambda=-1$. In this case, the topology does not change as
one goes from small to large $h^2$. All static solutions are
bounded. The admissible regions are completely filled by the three
JEF solutions, ${\rm cn}$, ${\rm dn}$, and staggered ${\rm dn}$
(marked as ${\rm dn}^*$). Several static solutions generated by
the map of Eq. (\ref{Two_point_map}) for $h=1.3$, panel (a), are
presented in Fig. \ref{Figure5}.} \label{Figure4}.
\end{figure}

The inadmissible regions are shown by the shaded areas in Fig.
\ref{Figure4} for (a) $h=1.3$ and (b) $h=4$. In this case, the
topology does not change as one goes from small to large $h^2$.
All static solutions are bounded. The admissible regions are
completely filled by the three JEF solutions, ${\rm cn}$, ${\rm
dn}$, and staggered ${\rm dn}$, marked as ${\rm dn}^*$ (see Sec.
\ref{Jacobi}).

\begin{figure}
\includegraphics{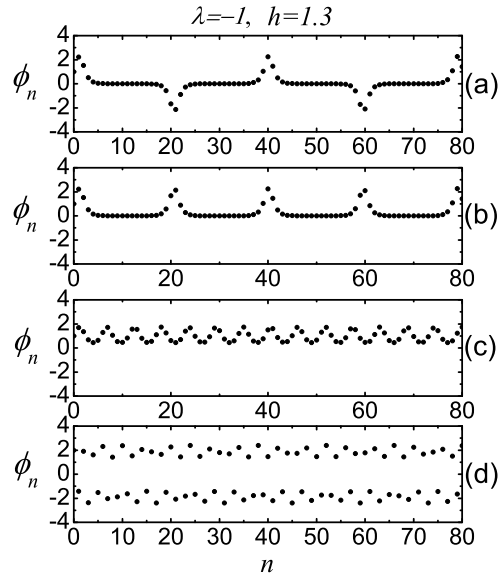}
\caption{Static solutions generated by the map of Eq.
(\ref{Two_point_map}) for $\lambda=-1$, $h=1.3$ [see panel (a) of
Fig. \ref{Figure4}] for different values of the integration
constants, $K$ and $\phi_0$. In (a), $K=-0.5-10^{-8}$, $\phi_0=1$; in
(b), $K=-0.5+10^{-8}$, $\phi_0=1$; in (c), $K=-0.2$, $\phi_0=1$; and in
(d), $K=5.3$, $\phi_0=2$. All the solutions are bounded. The solutions in (a)
and (b) are the ${\rm cn}$ and ${\rm dn}$ solutions close to the
hyperbolic function limit, respectively. The solution in (c) is the
${\rm dn}$, while in (d) we show the staggered ${\rm dn}$ solution.}
\label{Figure5}
\end{figure}

Several static solutions generated by the map of Eq.
(\ref{Two_point_map}) for $h=1.3$ [refer to panel (a) of Fig.
\ref{Figure4}] are presented in Fig. \ref{Figure5}. In (a),
$K=-0.5-10^{-8}$, $\phi_0=1$; in (b), $K=-0.5+10^{-8}$,
$\phi_0=1$; in (c), $K=-0.2$, $\phi_0=1$; and in (d), $K=5.3$,
$\phi_0=2$. The solutions in (a) and (b) are the ${\rm cn}$ and
${\rm dn}$ solutions close to the hyperbolic function limit,
respectively. These solutions can be regarded as a lattice of
pulses, Eq. (\ref{pulse}). Solutions shown in (c) and (d) are
${\rm dn}$ and staggered ${\rm dn}$ solutions, respectively.

\subsubsection{Case $\Lambda=2$} \label{Lambda2}

In this case, the two-point map of Eq. (\ref{Two_point_map})
reduces to
\begin{eqnarray} \label{ReducedMap}
   \phi _n^2 \left( {\phi _n^2 \phi _{n + 1}^2  - 2\tilde K\phi _n
   \phi _{n + 1}  + \tilde K} \right) = 0,
\end{eqnarray}
from which one has the trivial solution $\phi_n=0$ and, for
$\phi_n\neq 0$, one finds
\begin{eqnarray} \label{ReducedMap2}
   \phi _{n + 1}  = \frac{1}{{\phi _n }}\left( {\tilde K \pm \sqrt
{\tilde K(\tilde K-1)} } \right).
\end{eqnarray}
This map defines the two-parameter space of solutions. Admissible
regions of the $(K,\phi_0)$ plane correspond to $\tilde{K}\le 0$
and $\tilde{K}\ge 1$. Due to the symmetry of Eq.
(\ref{ReducedMap}), one can interchange $\phi_n$ and $\phi_{n+1}$
in Eq. (\ref{ReducedMap2}). From the two different roots of Eq.
(\ref{ReducedMap2}) one must take $\phi_{n+1}$ such that the three
point problem Eq. (\ref{Barashenkov}) is satisfied for $\phi_n$;
this would guarantee that one keeps moving along the same solution
branch. It is sufficient to take $\phi_{n+1}$ different from
$\phi_{n-1}$, if the roots are different.

It is convenient to take the integration constant in the form
\begin{eqnarray} \label{NewConstant}
   \tilde K =\frac{(1+a)^2}{4a},
\end{eqnarray}
and consider $-1 \le a \le 1$ with $a \neq 0$. With this choice,
for $-1 \le a < 0$ we have $-\infty < \tilde K \le 0$, while for
$0 < a \le 1$ we have $1 \le \tilde K < +\infty$.

Taking for the initial value $\phi_0 = b$, where $b \neq 0$ is
positive or negative real number, one can obtain from the map of Eq.
(\ref{ReducedMap2}) the following sequence, that satisfies
both Eq. (\ref{Barashenkov}) and Eq. (\ref{ReducedMap}),
\begin{eqnarray} \label{sol2}
   \phi_n: \left(...,b,\frac{(1+a)}{2b},\frac{b}{a},\frac{(1+a)a}{2b},\frac{b}{a^2},
   \frac{(1+a)a^2}{2b},...\right).
\end{eqnarray}
Note that for one of the bounds of the existence region,
$\tilde{K}=1$ ($a=1$), solution Eq. (\ref{sol2}) gives the
following family of two-site periodic solution
\begin{eqnarray} \label{sol1}
   \phi_n: \left(...,b,\frac{1}{b},b,\frac{1}{b},...\right).
\end{eqnarray}
On the other hand, the other bound of $\tilde{K}=0$ ($a=-1$) gives
from Eq. (\ref{sol2}) the following four-site periodic solution
\begin{eqnarray} \label{sol0}
   \phi_n: \left(...,b,0,-b,0,b,0,...\right).
\end{eqnarray}

It is worth pointing out that for the MC model (\ref{mc}) at
$\Lambda=2$, the two-point map has the form
\begin{eqnarray} \label{mcTwoPoint}
   (1-\phi_n^2)[\phi_{n+1}^2(1-\phi_n^2) - (1-\phi_n^2) + \tilde C]=0,
\end{eqnarray}
and, for $\phi_n \neq \pm1$, we obtain
\begin{eqnarray} \label{mcTwoPointSolved}
   \phi_{n+1}=\pm\sqrt{1-\frac{\tilde C}{1-\phi_n^2}}.
\end{eqnarray}
Based on this expression, starting from any admissible value of
$\phi_0$ and for any chosen $\tilde C$, one can construct a
solution iteratively. The topology of the admissible regions in this
case is not as simple as in the case of model Eq.
(\ref{Barashenkov}) at $\Lambda=2$.

A four-site periodic solution for $\phi_n$ in this case is given by
\begin{eqnarray} \label{solmc2}
  &&\phi_n: (...,a,b,-a,-b,a,b,-a,-b,a,...)\,, \nonumber \\
  &&\tilde{C}=(1-a^2)(1-b^2)\,,
\end{eqnarray}
where $a,b$ are any real numbers, positive or negative. In the
special case of $\tilde{C}=0$, apart from the above solution,
another solution for the
sequence $\phi_n$ is the arbitrary sequence of $\pm 1$ with $0$
put at random between $-1$ and +1 or between +1 and $-1$.

\subsection{Static Jacobi elliptic function solutions} \label{Jacobi}

Motivated by the results of the previous section, let us now
proceed to analytically identify several static JEF solutions to
Eq. (\ref{Barashenkov}). The solutions can be derived from the
three-point static problem of Eq. (\ref{Barashenkov}) with $\ddot{
\phi}_n=0$. In this case the use of the JEF identities listed in
the Appendix II is particularly helpful. Alternatively, solutions
can be derived from the DFI of Eq. (\ref{Two_point_map}). In this
case, one also obtains the relation between the integration
constant $K$ and the parameters of the JEF solution thus
establishing a link between the solutions obtained by two
different approaches.

The results will be presented in the following way. We will first
describe the general form of the derived JEF solutions, then give
the parameters for the particular ones, and finally, we will
characterize the solutions and describe the ranges of parameters
where different solutions exist.

The general form of the solutions is
\begin{eqnarray} \label{EllipticDiscrete}
   \phi_n&=&\pm SA {\rm sn}^p(Z,m) {\rm cn}^q(Z,m) {\rm dn}^r(Z,m) ,
   \nonumber \\
   Z&=&\beta h \left(n+x_0\right),
\end{eqnarray}
where ${\rm sn}$, ${\rm cn}$, and ${\rm dn}$ are the Jacobi
elliptic functions, $0 \le m \le 1$ is the modulus of JEF, $A$ and
$\beta$ are the parameters of the solution, and $x_0$ is the
arbitrary initial position. For non-staggered solutions $S=1$,
while for staggered ones $S=(-1)^n$. Finally the integers $p,q,r$
specify a particular form of the solution.

For the sake of brevity we introduce the following notations
\begin{eqnarray} \label{scd}
   s = {\rm{sn}}(\beta h,m), \quad c = {\rm{cn}}(\beta h,m),
   \quad d = {\rm{dn}}(\beta h,m).
\end{eqnarray}

Particular solutions have the following form and are characterized
by the following parameters:

The ${\rm sn}$ solution, $(p,q,r)=(1,0,0)$,
\begin{eqnarray} \label{Elliptic_sn}
    && A^2=\frac{(2-\Lambda)ms^2}{\Lambda}\,,~~
   A^2 =\frac{2ms^2cd}{\Lambda(1-ms^4)}\,, \nonumber \\
    && \tilde K = \frac{A^4}{m}, \quad S=1.
\end{eqnarray}

The ${\rm cn}$ solution, $(p,q,r)=(0,1,0)$,
\begin{eqnarray} \label{Elliptic_cn}
   &&A^2=\frac{(\Lambda-2)ms^2}{\Lambda d^2}\,,~~
   A^2= -\frac{2ms^2c}{\Lambda(d^2-ms^2c^2)}\,, \nonumber \\
   &&\tilde{K}=-(1-m)\frac{A^4}{m}, \quad S=1.
\end{eqnarray}

The ${\rm dn}$ solution, $(p,q,r)=(0,0,1)$,
\begin{eqnarray} \label{Elliptic_dn}
   &&A^2=\frac{(\Lambda-2)s^2}{\Lambda c^2}\,,~~
  A^2 = -\frac{2s^2d}{\Lambda(c^2-s^2d^2)}\,, \nonumber \\
   &&\tilde{K}=(1-m)A^4, \quad S=1.
\end{eqnarray}

The staggered ${\rm sn}^*$ solution, $(p,q,r)=(1,0,0)$,
\begin{eqnarray} \label{Elliptic_ssn}
   &&A^2=\frac{(2-\Lambda)ms^2}{\Lambda}\,,~~
  A^2 =-\frac{2ms^2cd}{\Lambda(1-ms^4)}\,, \nonumber \\
   &&\tilde{K}=\frac{A^4}{m}, \quad S=(-1)^n.
\end{eqnarray}

The staggered ${\rm cn}^*$ solution, $(p,q,r)=(0,1,0)$,
\begin{eqnarray} \label{Elliptic_scn}
   &&A^2=\frac{(\Lambda-2)ms^2}{\Lambda d^2}\,,~~
  A^2 =\frac{2ms^2c}{\Lambda(d^2-ms^2c^2)}\,, \nonumber \\
   &&\tilde{K}=-(1-m)\frac{A^4}{m}, \quad S=(-1)^n.
\end{eqnarray}

The staggered $\rm{dn}^*$ solution, $(p,q,r)=(0,0,1)$,
\begin{eqnarray} \label{Elliptic_sdn}
   &&A^2=\frac{(\Lambda-2)s^2}{\Lambda c^2}\,,~~ A^2=
   \frac{2s^2d}{c^2-s^2d^2}\,, \nonumber \\
   &&\tilde{K}=(1-m)A^4, \quad S=(-1)^n.
\end{eqnarray}

The $\rm{1/sn}$ solution, $(p,q,r)=(-1,0,0)$,
\begin{eqnarray} \label{Elliptic_inv_sn}
   &&A^2 = \frac{2s^2cd}{\Lambda (1-ms^4}\,,~~
   A^2 =\frac{(2 - \Lambda)s^2}{ \Lambda} , \nonumber \\
   && \tilde K = A^4 m, \quad S=1.
\end{eqnarray}

The $\rm{1/cn}$ solution, $(p,q,r)=(0,-1,0)$,
\begin{eqnarray} \label{Elliptic_inv_cn}
   &&A^2 = \frac{2(1-m)cs^2}{\Lambda(d^2-ms^2c^2}\,,~~
  A^2 = \frac{(2 - \Lambda)(1 - m)s^2}{\Lambda d^2},\nonumber \\
   &&\tilde K = \frac{- mA^4}{1 - m}, \quad S=1.
\end{eqnarray}

The staggered $({\rm sn/cn})^*$ solution, $(p,q,r)=(1,-1,0)$,
\begin{eqnarray} \label{Elliptic_sncn_stagg}
   &&A^2  = \frac{(1-m)(2-\Lambda)s^2}{\Lambda c^2}\,,~~
  A^2 = -\frac{2(1-m)s^2d}{\Lambda(c^2-s^2d^2) },\nonumber \\
   &&\tilde K = \frac{A^4}{1-m}, \quad S=(-1)^n.
\end{eqnarray}

The ${\rm sndn/cn}$ solution, $(p,q,r)=(1,-1,1)$,
\begin{eqnarray} \label{Elliptic_sndncn}
   &&A^2  = \frac{2 s^2d^2(1-2ms^2 +ms^4)}
   {\Lambda (c^4 - s^4 d^4)}\,,~~
    A^2 = \frac{(2 - \Lambda)s^2 d^2}{\Lambda c^2},\nonumber \\
   &&\tilde K = A^4, \quad S=1.
\end{eqnarray}

The staggered $({\rm sndn/cn})^*$ solution, $(p,q,r)=(1,-1,1)$,
\begin{eqnarray} \label{Elliptic_sndncn_stagg}
   &&A^2  = -\frac{2 s^2d^2(1-2ms^2 +ms^4)}
   {\Lambda (c^4 - s^4 d^4)}\,,~~
    A^2 = \frac{(2 - \Lambda)s^2 d^2}{\Lambda c^2},\nonumber \\
   &&\tilde K = A^4, \quad S=(-1)^n.
\end{eqnarray}

It is worth making the following remarks:

To specify a particular solution, for given model parameters $h^2$
and $\lambda$, and for chosen $m$, one has to solve the first two
equations entering Eqs.
(\ref{Elliptic_sn})-(\ref{Elliptic_sndncn_stagg}) in order to find
the parameter $\beta$ and then the amplitude $A$. If the equation for
$\beta$ has no solutions in the range of $0< \beta h <
2K(m)$, this implies that for chosen $h^2,\,\,\lambda$, and $m$ the
corresponding JEF solution does not exist. Some of the roots for
$\beta$ can give imaginary amplitude $A$ and we reject such solutions.
On the other hand, if one or more roots for
$\beta$ in the range of $0< \beta h < 2K(m)$ are found, then a JEF
solution can be constructed for each root provided that the
corresponding amplitude $A$ is real. In Table \ref{Table1}, for
different JEF solutions, we summarize the location of roots for
the parameter $\beta$, corresponding to real amplitude $A$ and
give the range of parameter $\Lambda$ where the solutions exist.
For comparison, similar results for bounded JEF solutions to the
MC model of Eq. (\ref{mc}) are given in Table \ref{Table2}.

\begin{table}
\caption{Range of parameter $\Lambda$ and location of the root for
parameter $\beta$ corresponding to real amplitude $A$ for JEF
solutions to Eq. (\ref{Barashenkov})\label{Table1}}
\begin{tabular}{|l|l|l|}  \hline
  Solution    & Root for $\beta$ within & Root for $\beta$ within \\
              & $0<\beta h<K(m)$        & $K(m)<\beta h<2K(m)$ \\ \hline
  sn, Eq. (\ref{Elliptic_sn})    & $0<\Lambda<2$ & no solution \\ \hline
  cn, Eq. (\ref{Elliptic_cn})    & $\Lambda<0$   & $\Lambda>2$ \\ \hline
  dn, Eq. (\ref{Elliptic_dn})    & $\Lambda<0$   & $\Lambda>2$ \\ \hline
  sn$^*$, Eq. (\ref{Elliptic_ssn})    & no solution & $0 <\Lambda<2$ \\ \hline
  cn$^*$, Eq. (\ref{Elliptic_scn})    & $\Lambda>2$   & $\Lambda<0$ \\ \hline
  dn$^*$, Eq. (\ref{Elliptic_sdn})    & $\Lambda>2$   & $\Lambda<0$ \\ \hline
  1/sn, Eq. (\ref{Elliptic_inv_sn})    & $0 <\Lambda<2$   & no solution \\ \hline
  1/cn, Eq. (\ref{Elliptic_inv_cn})    & $0 <\Lambda<2$   & no solution \\ \hline
  (sn/cn)$^*$,  & $0 <\Lambda<2$ & $0 <\Lambda<2$ \\
  Eq. (\ref{Elliptic_sncn_stagg}) &                  &                 \\ \hline
  sndn/cn,                    & $0 <\Lambda<1$ & $0 <\Lambda<1$ \\
  Eq. (\ref{Elliptic_sndncn}) &                &              \\ \hline
  (sndn/cn)$^*$,   & $0 <\Lambda<2$ & $0 <\Lambda<2$ \\
  Eq. (\ref{Elliptic_sndncn_stagg}) &     &     \\ \hline
\end{tabular}
\end{table}

\begin{table}
\caption{Same as in Table \ref{Table1} but for the bounded JEF
solutions to the MC model Eq. (\ref{mc})\label{Table2}}
\begin{tabular}{|l|l|l|}  \hline
  Solution    & Root for $\beta$ within & Root for $\beta$ within \\
              & $0<\beta h<K(m)$        & $K(m)<\beta h<2K(m)$ \\ \hline
  sn    & $0<\Lambda<2$ & $\Lambda>2$ \\ \hline
  cn    & $\Lambda<0$   & no solution \\ \hline
  dn    & $\Lambda<0$   & no solution \\ \hline
  sn$^*$    & $\Lambda>2$ & $0 <\Lambda<2$ \\ \hline
  cn$^*$    & no solution   & $\Lambda<0$ \\ \hline
  dn$^*$    & no solution   & $\Lambda<0$ \\ \hline
\end{tabular}
\end{table}

In each of Eqs. (\ref{Elliptic_sn})-(\ref{Elliptic_sndncn_stagg})
we give the corresponding expressions for $\tilde K$ in terms of
the parameters of JEF solutions. These expressions link the JEF
solutions to the nonlinear map Eq. (\ref{Two_point_map}), so that
for each JEF solution one can find the corresponding portion of
the $(K,\phi_0)$ plane. For example, the ${\rm sn}$ solution
exists for $0<\Lambda<2$ (see Table \ref{Table1}) and this case is
shown in Fig. \ref{Figure1}. Then, the ${\rm sn}$ solution having
modulus $m$ and amplitude $A$, in the $(K,\phi_0)$ plane will have
abscissa $K$ that can be found for given $\tilde K$ from Eq.
(\ref{KFarbtilde}) and the ordinate will vary in the range $-A \le
\phi_0 \le A$. Considering different values of $0 \le m \le 1$,
one can see that the ${\rm sn}$ solution occupies the portions
marked with ``sn" in Fig. \ref{Figure1}.

The solutions presented by Eqs.
(\ref{Elliptic_sn})-(\ref{Elliptic_sdn}) are bounded while the
other solutions are unbounded.

The ${\rm sn}$ solution and its staggered form ${\rm sn}^*$ are in
fact equivalent in the sense that any sequence of $\phi_n$
produced by one of them can also be produced by the other. As a
result, ${\rm sn}$ and ${\rm sn}^*$ solutions occupy the same
portion of the $(K,\phi_0)$ plane (see Fig. \ref{Figure1}). The
same can be said about the pair of solutions ${\rm cn}$ and ${\rm
cn}^*$. However, the ${\rm dn}$ solution is essentially different
from its staggered form, ${\rm dn}^*$, and also the ${\rm
sndn/cn}$ solution is essentially different from its staggered
form, $({\rm sndn/cn})^*$.

\subsection{Hyperbolic function static solutions} \label{Hyperbolic}

In the limit of $m\rightarrow 1$ the $\rm{sn}$ solution given by
Eqs. (\ref{EllipticDiscrete}) and (\ref{Elliptic_sn}) reduces to
the kink (antikink) solution
\begin{eqnarray} \label{kink}
   \phi_n = \pm \tanh[\beta h ( n + x_0 )],
\end{eqnarray}
where $x_0$ is an arbitrary shift, $\beta$ satisfies
\begin{eqnarray} \label{tanh_beta}
   \tanh^2 ( \beta h ) = \frac{ \Lambda}{2 - \Lambda},
\end{eqnarray}
and $\tilde{K}=1$, i.e., $K=0$.

In the limit of $m\rightarrow 1$, both the $\rm{cn}$ solution and
the $\rm{dn}$ solution with parameters given by Eq.
(\ref{Elliptic_cn}) and Eq. (\ref{Elliptic_dn}), respectively,
reduce to the same pulse solution
\begin{eqnarray} \label{pulse}
   \phi_n = \pm A{\rm sech}[\beta h ( n + x_0 )],
\end{eqnarray}
where $x_0$ is an arbitrary shift, $\beta$ and $A$ can be found from
\begin{eqnarray} \label{sech_beta}
   {\rm{cosh}}(\beta h) &=& \frac{2 - \Lambda}{{2}}, \nonumber \\
  \Lambda A^2 &=& (\Lambda-2)\sinh^2(\beta h),
\end{eqnarray}
and $\tilde{K}=0$, i.e., for $\lambda=-1$, $K=-1/2$.

We now turn to the case of staggered hyperbolic function
solutions. It is easily checked from Eq. (\ref{Elliptic_ssn}) that
the TI model of Eq. (\ref{Barashenkov}) does not support the
staggered kink solution (even though it supports the $\rm{sn}^*$
solution). However, it supports the staggered pulse solution
\begin{eqnarray} \label{pulse_stagg}
   \phi_n = \pm (-1)^{n} A{\rm sech}[\beta h ( n + x_0 )],
\end{eqnarray}
where $x_0$ is an arbitrary shift, $\beta$ and $A$ can be found from
\begin{eqnarray} \label{sech_beta_stagg}
   {\rm{cosh}}(\beta h)&=&\frac{{\Lambda  - 2}}{2},\nonumber \\
   \Lambda A^2&=&\left( {\Lambda  - 2} \right){\rm{sinh}}^{\rm{2}}
   (\beta h).
\end{eqnarray}
and $\tilde{K}=0$, i.e., for $\lambda=1$, $K=1/2$.

In Table \ref{Table3} we present the ranges of model parameter
$\Lambda$ where the hyperbolic function solutions exist in the
model of Eq. (\ref{Barashenkov}) and, for comparison, in Table
\ref{Table4} we give similar results for the MC model of Eq.
(\ref{mc}).

\begin{table}
\caption{Range of parameter $\Lambda$ for hyperbolic solutions to
Eq. (\ref{Barashenkov}) \label{Table3}}
\begin{tabular}{|l|l|}  \hline
  Solution    & Exists for   \\ \hline
  Kink, Eq. (\ref{kink})    & $0 < \Lambda < 1$   \\ \hline
  Pulse, Eq. (\ref{pulse})    & $ \Lambda < 0$   \\ \hline
  Staggered kink    & no solution   \\ \hline
  Staggered pulse, Eq. (\ref{pulse_stagg})    & $ \Lambda > 4$   \\ \hline
\end{tabular}
\end{table}

\begin{table}
\caption{Same as in Table \ref{Table3} but for MC model Eq.
(\ref{mc}) \label{Table4}}
\begin{tabular}{|l|l|}  \hline
  Solution    & Exists for   \\ \hline
  Kink    & $0 < \Lambda < 2$   \\ \hline
  Pulse    & $ \Lambda < 0$   \\ \hline
  Staggered kink    & $ \Lambda > 2$   \\ \hline
  Staggered pulse    & no solution   \\ \hline
\end{tabular}
\end{table}

\section{Linear stability and dynamics for bounded solutions for $h^2<2$}
\label{Numerics}

\subsection{Linear stability} \label{LinearStability1}

Introducing the ansatz $\phi_n(t)=\phi_n^0+\varepsilon_n(t)$
(where $\phi_n^0$ is an equilibrium solution and
$\varepsilon_n(t)$ is a small perturbation), we linearize Eq.
(\ref{Barashenkov}) with respect to $\varepsilon_n$, in order
to examine the linear stability of the above-established
solutions.  In this way, we obtain the
following equation:
\begin{eqnarray} \label{Linearized}
   \ddot{\varepsilon}_n=\frac{1}{h^2}
   (\varepsilon_{n-1}-2\varepsilon_{n}+\varepsilon_{n+1})
   + \lambda \varepsilon_{n} \nonumber \\
   -\lambda \phi_{n}^0\phi_{n+1}^0 \varepsilon_{n-1}
   -\lambda \phi_{n-1}^0\phi_{n+1}^0 \varepsilon_{n}
   -\lambda \phi_{n-1}^0\phi_{n}^0 \varepsilon_{n+1} .
\end{eqnarray}
For the small-amplitude phonons, $\varepsilon_{n}=\exp(i k n +i
\omega t)$, with frequency $\omega$ and wave number $k$, Eq.
(\ref{Linearized}) is reduced to the following dispersion
relation:
\begin{eqnarray} \label{SpecClassic1}
   \omega^2=\frac{4}{h^2}\sin^2
   \left( \frac{k}{2} \right) -\lambda +\lambda \phi_{n-1}^0\phi_{n+1}^0 \nonumber \\
   + \lambda \phi_{n}^0\phi_{n+1}^0 e^{-ik}
   + \lambda \phi_{n-1}^0\phi_{n}^0 e^{ik},
\end{eqnarray}
where it is assumed that $\phi_n^0$ is a uniform steady state
solution (or a zigzag solution; see below).
The corresponding eigenvalue problem has a non-symmetric matrix. This
fact is a manifestation of the non-Hamiltonian nature
of Eq. (\ref{Barashenkov}).

{\it Stability of vacuum solutions.} From Eq.
(\ref{SpecClassic1}), the spectrum of the vacuum solution
$\phi_n^0=\pm 1$ (at $\lambda=1$) is
\begin{equation}\label{SpecVacuum1}
   \omega^2=2+ \frac{4}{h^2}(1-h^2)\sin^2\left(\frac{k}{2}\right) .
\end{equation}
This solution is stable for $h^2 \le 2$.

The spectrum of the vacuum solution $\phi_n^0=0$ (at $\lambda=-1$) is
\begin{equation}\label{SpecVacuum2}
      \omega^2=1 + \frac{4}{h^2}\sin^2
   \left( \frac{k}{2} \right),
\end{equation}
and this solution is stable for any value of $h^2$.

\begin{figure}
\includegraphics{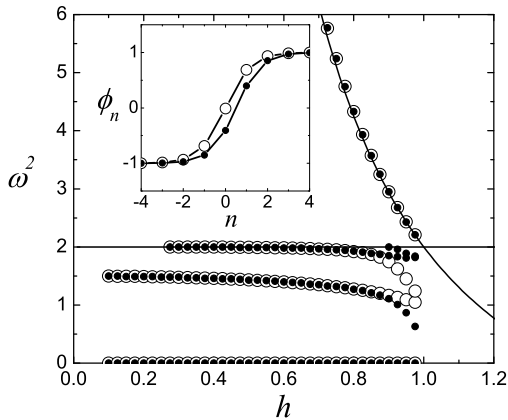}
\caption{Frequencies of the kink's internal modes for different
magnitudes of the discreteness parameter $h$. Results for the
on-site (circles) and inter-site (dots) kinks. The corresponding
kink profiles are shown in the inset for $h=0.8$. Two solid lines
show the borders of the spectrum of vacuum, Eq.
(\ref{SpecVacuum1}). The kink possesses the translational internal
mode with $\omega^2=0$ and also the localized modes with
frequencies below the phonon band and a localized mode with the
frequency above the phonon band. The kink is stable in the whole range
of its existence, $h^2<1$.} \label{Figure6}
\end{figure}

{\it Stability of kink solution.} As it can be seen from Eq.
(\ref{tanh_beta}), the kink solution Eq. (\ref{kink}) exists only
for $\lambda=1$ and $h^2<1$. The spectrum of lattice with a kink as
a function of lattice spacing $h$ is shown in Fig. \ref{Figure6}
for the on-site kink, $x_0=0$ (open circles), and inter-site kink,
$x_0=h/2$ (dots). The corresponding kink profiles are shown in the
inset for $h=0.8$. Two solid lines show the borders of the spectrum
of vacuum, Eq. (\ref{SpecVacuum1}). The kink possesses the
translational internal mode with $\omega^2=0$ and also the
localized modes with frequencies below the phonon band and a
localized mode with frequency above the phonon band. The kink
is stable in the whole range of existence, $h^2<1$. Notice also
that as the continuum limit $h \rightarrow 0$ is approached,
a number of internal modes that exist for larger $h$ disappear
into the phonon band of Eq.
(\ref{SpecVacuum1}); the only internal mode that persists
with $\omega \rightarrow \sqrt{3}$, as $h \rightarrow 0$
 is the well-known continuum $\phi^4$ model
internal mode (see e.g. \cite{PhysicaD} and references therein).

{\it Instability of the zigzag solution}. In the sinusoidal
function limit $(m=0)$, the staggered ${\rm dn}$ solution defined
by Eq. (\ref{EllipticDiscrete}) and Eq. (\ref{Elliptic_sdn})
reduces to the zigzag solution
\begin{eqnarray} \label{zigzag}
   \phi_n=(-1)^nA, \quad \quad
   A^2=\frac{\Lambda - 4}{\Lambda}.
\end{eqnarray}
Note that this solution is valid if $\Lambda >4$ or if $\Lambda <
0$. Substituting this solution into Eq. (\ref{SpecClassic1}) we
obtain the dispersion relation $\omega^2 h^2 = 4 - 2\Lambda +
4(\Lambda - 3)\sin^2(k/2)$. One can see that the spectrum of the
zigzag solution always contains imaginary frequencies in both
cases, $\Lambda >4$ and $\Lambda < 0$, and thus, the solution is
always unstable.

{\it Stability of solutions at $\Lambda=2$}. In this case (see
Sec. \ref{Lambda2}), we have two bounded solutions, Eq.
(\ref{sol1}) and Eq. (\ref{sol0}). For the two-periodic solution
Eq. (\ref{sol1}), the vibrational spectrum has two branches,
$(h^2\omega^2)_{1,2} = a^2 + 1/a^2 \pm \sqrt{(a^2 + 1/a^2)^2 -2
+2\cos k}$. Since $a^2 + 1/a^2 \ge 2$ for any $a$, $\omega^2$ is
always non-negative and the solution is always stable. Numerically
we found that the four-periodic solution Eq. (\ref{sol0}) is
always unstable. This can be justified by introducing a small
deviation $\varepsilon$ into the position of a particle with
$\phi=0$, while keeping all other particles in their equilibrium
positions. The right-hand side of Eq. (\ref{Barashenkov}) gives
the force acting on the particle equal to $2\varepsilon a^2$.
Since the sign of the force coincides with the sign of the
displacement, the solution is indeed unstable with respect to the
considered perturbation.

For the MC model Eq. (\ref{mc}) in the case of $\Lambda=2$, we
found that the solutions $...,-1,1,-1,1,...$ and
$...,-1,0,1,-1,0,1,...$ are unstable. This can be checked by
introducing a small displacement into position of a particle with
$\phi=1$, keeping other particles at their equilibrium positions.
We then calculate the increment of force acting on this particle
and find that it has the same sign as the displacement, which
proves the instability with respect to this type of perturbation.

Calculating the increment of force acting on a particle with
unperturbed amplitude $\phi=1$ in response to a small displacement
$\varepsilon$ of this particle, we found that it has the same sign
as the displacement. Thus, these two solutions are unstable and
this fact was also confirmed numerically.

{\it Stability of other solutions}. We also examined numerically
the stability of other ones among the above-found solutions. We
studied bounded, periodic Jacobi elliptic function solutions
depicted in Fig. \ref{Figure1} and Fig. \ref{Figure4}(a), i.e.,
the $\rm sn$, $\rm cn$, $\rm dn$, and staggered $\rm dn$
solutions. The length of the chain was compatible with the period
of the solution and it contained a few periods. The eigenvalue
problem obtained by replacing the left-hand side of Eq.
(\ref{Linearized}) with $-\omega^2\varepsilon_n$ was solved
numerically under periodic boundary conditions. The ${\rm sn}$,
${\rm cn}$, $\rm dn$, staggered $\rm dn$, and the pulse solutions
were found to be generically unstable, i.e., having negative
eigenfrequencies $\omega^2$ in their numerical spectra.

It is important to note that all the solutions possess the
zero-frequency translational eigenmode due to the TI nature of the
lattice of Eq. (\ref{Barashenkov}). As an overview of the results
of the stability analysis, we find that, among the bounded solutions,
only the kink solution, the vacuum solutions, and the solution of Eq.
(\ref{sol1}) ($\Lambda=2$ case) can be stable. Note that exactly
the same conclusion was also reached in the case of the MC model
\cite{DKYF_PRE2006} in case $\Lambda \ne 2$.

\subsection{Dynamics of unstable solutions} \label{Dynamics}

\begin{figure}
\includegraphics{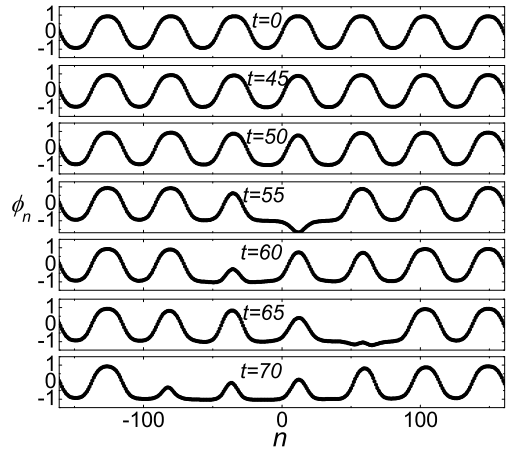}
\caption{Time evolution of the ${\rm sn}$ solution perturbed at
$t=0$ by shifting the position of $n=0$ particle by $-10^{-2}$. The model
parameters are $F=\lambda=1$, $h=0.25$ and the parameters of the
${\rm sn}$ solution are $m\simeq 0.777$, $x_0=0$.}
\label{Figure7}
\end{figure}

\begin{figure}
\includegraphics{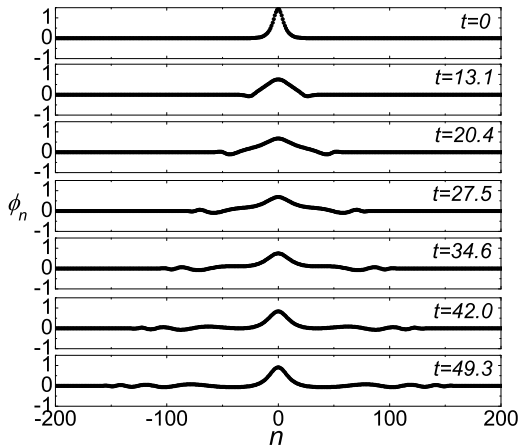}
\caption{Time evolution of the pulse solution perturbed at $t=0$
by shifting the position of $n=0$ particle by $-10^{-4}$.  The pulse is
shown at the moment of maximum displacement of the central
particle. The period of oscillation of the pulse is roughly equal to
$7$. The model parameters are $F=\lambda=-1$, $h=0.25$ and the pulse
is placed at $x_0=0$.} \label{Figure8}
\end{figure}

\begin{figure}
\includegraphics{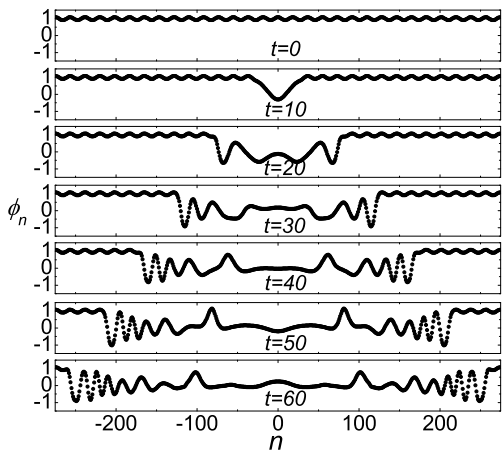}
\caption{Time evolution of the ${\rm dn}$ solution perturbed at
$t=0$ by shifting the position of the $n=0$ particle by $-10^{-4}$. The model
parameters are $F=\lambda=-1$, $h=0.25$ and the parameters of the
${\rm dn}$ solution are $m=0.3$, $x_0=0$.} \label{Figure9}
\end{figure}

In the figures \ref{Figure7}-\ref{Figure9} we
give examples of the dynamical behavior of the unstable static
solutions perturbed at $t=0$ by adding a small displacement to
the $0$-th particle.

In Fig. \ref{Figure7} the ${\rm sn}$ solution was perturbed at
$t=0$ by shifting the position of $n=0$ particle by $-10^{-2}$. The model
parameters are chosen as $\lambda=1$, $h=0.25$ and the parameters of the
${\rm sn}$ solution are $m\simeq 0.777$, $x_0=0$. Up to the time
$t\approx 45$ the long-wave mode of deviation from equilibrium,
growing exponentially with time, remains indistinguishable in the
scale of the figure. After that, the ${\rm sn}$ solution gradually
slides down into the potential well at $\phi=-1$ and a set of
breather-like oscillating pulses is formed. The oscillations are
accompanied by phonon radiation and their amplitude decreases in time.
The instability picture presented in Fig. \ref{Figure7} can also
be interpreted in terms of the kink-antikink interactions. Such
interactions are attractive and the chain of mutually
attractive quasi-particles, as it is well-known, is unstable and
cannot remain equally spaced in the presence of small
perturbations.

Similar behavior is demonstrated by the unstable pulse perturbed
at $t=0$ by shifting the position of the $n=0$ particle by $-10^{-4}$ (see
Fig. \ref{Figure8}). The pulse slides down into the potential well
at $\phi=0$ and oscillates by radiating energy. The pulse is shown at
the moment of maximum displacement of the central particle. The period
of oscillation of the pulse is roughly equal to $7$. The model parameters
are $\lambda=-1$, $h=0.25$ and the pulse is placed at $x_0=0$.
We note in passing that the instability of the single pulse has been
demonstrated quite generally for discrete TI settings in the case of the
$\phi^4$ equation in \cite{DKYF_PRE2006}; on the basis of that proof,
we expect the individual pulses (as well as solutions consisting of
concatenations of such pulses) to be unstable in the present model
as well.

In Fig. \ref{Figure9} the time evolution of the unstable ${\rm
dn}$ solution perturbed at $t=0$ by shifting the position of the $n=0$
particle by $-10^{-4}$ is presented. The model parameters are
$\lambda=-1$, $h=0.25$ and the parameters of the ${\rm dn}$
solution are $m=0.3$, $x_0=0$. The ${\rm dn}$ solution has
particles situated near the top of the potential barrier at
$\phi=1$ and, being perturbed as described above, the particles,
one after another, slide down into the potential well at $\phi=0$
and oscillate near the potential well.

Note that the perturbation of the opposite sign (outward of the
potential well at $\phi=0$) would cause the divergence of the
pulse and the ${\rm dn}$ solutions shown in the top panels of Fig.
\ref{Figure8} and Fig. \ref{Figure9}, respectively.
We note that qualitatively similar behavior was also observed
in the case of the MC model \cite{DKYF_PRE2006}.

\section{Discussion and conclusions} \label{Conclusions}

For a non-standard, translationally invariant discrete lattice,
namely the discretization of the $\phi^4$ model of Eq.
(\ref{Barashenkov}), we have reported a number of important
developments. We have explicitly identified the momentum
conservation law given by Eq. (\ref{mom2}), produced a set of
static solutions in terms of the Jacobi elliptic functions,
including the staggered ones, see Sec. \ref{Jacobi} and their
hyperbolic limit. We have fully analysed the existence of
such solutions, as well as detailed their linear stability. In
Appendix I, we have demonstrated how, using a JEF solution of Eq.
(\ref{Barashenkov}), one can construct the nonlinear map (or DFI)
of Eq. (\ref{Two_point_map}). Importantly, we have proposed
this as an illustrative
example of how to formulate the discrete first integral approach,
how to link it to the Jacobi elliptic function solutions, how to
construct the complete space of solutions of the static problem,
and identify the associated conservation law for such discrete,
translationally invariant models.

An important feature of our approach is that
from the nonlinear map Eq. (\ref{Two_point_map}), {\it any} static
solution of Eq. (\ref{Barashenkov}) can be obtained recurrently,
solving at each step a simple algebraic problem, in our case,
a quadratic equation. Admissible values of the integration constants
$K$ and $\phi_0$ have been plotted for positive and negative
$\lambda$ and for different values of the discreteness parameter
$h^2$, see Figs. \ref{Figure1}, \ref{Figure3}, and \ref{Figure4}.
Any point inside the admissible region corresponds to the static
solutions of Eq. (\ref{Barashenkov}).

We expect that the results presented herein for the discrete
$\phi^4$ equation can be applied to analogous discrete nonlinear
Schr{\"o}dinger type equations, similarly to what was discussed in
the recent works \cite{DNLSE1,DNLSE2,DNLSE3,DEPelinovsky}. Another
very natural extension of the present approach, detailing how to
obtain an analytical handle on solutions of such translationally
invariant models, would be to consider higher dimensional
settings. Very little is known even in 2+1-dimensional spatially
discrete models about analytical solutions, hence such
investigations would be a particularly interesting next step in
this direction of research.

\section*{Appendix I} \label{AppendixI}

In this Appendix we show how the two-point maps can be constructed
directly from the known
exact solutions. For example, for the $\tanh$ solution Eq.
(\ref{kink}) one can find $\phi_{n-1}$ or $\phi_{n+1}$ using the
identity $\tanh(x+y)=(\tanh x+\tanh y)/(1 + \tanh x \tanh y)$ and
then, making use of Eq. (\ref{kink}) and Eq. (\ref{tanh_beta})
write down the two-point map that generates the $\tanh$ solution
\begin{eqnarray} \label{Two_point_tanh}
   \phi_{n\pm 1} = \frac{\phi_n \pm T}{1 \pm T \phi_n},
   \quad T=\sqrt{\frac{ \Lambda}{2 - \Lambda}}.
\end{eqnarray}
This map does not involve the integration constant because for the
$\tanh$ solution we have already set $m=1$. The initial value can be
chosen arbitrarily from the range $-1<\phi_0<1$.

Similarly, for the ${\rm sn}$ solution, Eq.
(\ref{EllipticDiscrete}) and Eq. (\ref{Elliptic_sn}), using the
formula for ${\rm sn}(x+y)$, we can write
\begin{eqnarray} \label{Two_point_sn1}
   \phi _{n \pm 1}  = A\frac{cdS \pm sCD}
   {1 - ms^2 S^2},
\end{eqnarray}
where we have denoted
\begin{eqnarray} \label{SSCCDD}
   S = {\rm{sn}}[\beta h ( n + x_0 ),m], \nonumber \\
   C = {\rm{cn}}[\beta h ( n + x_0 ),m], \nonumber \\
   D = {\rm{dn}}[\beta h ( n + x_0 ),m],
\end{eqnarray}
and also used the notations of Eq. (\ref{scd}). According to Eq.
(\ref{Elliptic_sn}) we have $\phi_n=AS$ and, using the identity
for the Jacobi elliptic functions $C^2D^2 = 1-(1+m)S^2 +mS^4$, we
rewrite Eq. (\ref{Two_point_sn1}) in the form of the two-point map
\begin{eqnarray} \label{Two_point_sn2}
   \phi _{n \pm 1}  = \frac{{Acd\phi _n  \pm s\sqrt { A^4  -
   (1 + m)A^2 \phi _n^2  + m\phi _n^4 } }}{{A
   - ms^2 \phi _n^2 /A}}.
\end{eqnarray}
The integration constant here is $0 \le m \le 1$. As soon as $m$
is specified, one can find $\beta$ and $A$ from Eq.
(\ref{Elliptic_sn}) and then $s$, $c$, $d$ from Eq.
(\ref{scd}), thus defining the map Eq. (\ref{Two_point_sn2}). The
corresponding ${\rm sn}$ solution can be constructed by iterating
Eq. (\ref{Two_point_sn2}) starting from any initial value taken
from $-A \le \phi_0 \le A$.

By analogy, two-point maps can be constructed for any solution
given in Sec. \ref{Jacobi} and Sec. \ref{Hyperbolic}. However,
each of the maps constructed in this way is parameterized by the
modulus $m$ and can be applied only to the corresponding solution.
In our case, it is possible to substitute $m$ with a universal
integration constant so that the resulting two-point map can
generate all static solutions of Eq. (\ref{Barashenkov}). The
universal integration constant can be taken, e.g., as
\begin{equation} \label{AII1}
   K = \frac{\lambda}{2}\left( 1 - \frac{A^4}{m} \right) ,
\end{equation}
then we have
\begin{equation} \label{AII2}
   \tilde{K} = 1 - \frac{2K}{\lambda} = \frac{A^4}{m}.
\end{equation}
From Eq. (\ref{Elliptic_sn}) one can derive
\begin{equation} \label{AII3}
   cd = Z, \quad \quad
   \frac{{\left( {1 + m} \right)A^2 }}{m} = X,
\end{equation}
where $Z$ and $X$ are given by Eq. (\ref{Z}) and Eq. (\ref{X}),
respectively. With the help of Eq. (\ref{Elliptic_sn}) and Eqs.
(\ref{AII1})-(\ref{AII3}) we transform Eq. (\ref{Two_point_sn2})
to the form of Eq. (\ref{Two_point_map}) which does not include
explicitly the parameters of the ${\rm sn}$ solution.

\section*{Appendix II} \label{AppendixII}

In the derivation of the Jacobi elliptic function solutions
presented in Sec. \ref{Jacobi}, the following identities for Jacobi
elliptic functions have been used \cite{kls}

\begin{eqnarray} \label{Identity1}
   m{\rm sn}(x+a){\rm sn}(x){\rm sn}(x-a) \nonumber \\
   ={\rm ns}(a){\rm ns}(2a)[{\rm sn}(x+a)+{\rm sn}(x-a)] \nonumber \\
   -{\rm ns}^2(a){\rm sn}(x),
\end{eqnarray}

\begin{eqnarray} \label{Identity2}
   m{\rm cn}(x+a){\rm cn}(x){\rm cn}(x-a) \nonumber \\
   =-{\rm ds}(a){\rm ds}(2a)[{\rm cn}(x+a)+{\rm cn}(x-a)] \nonumber \\
   +{\rm ds}^2(a){\rm cn}(x),
\end{eqnarray}

\begin{eqnarray} \label{Identity3}
   m{\rm dn}(x+a){\rm dn}(x){\rm dn}(x-a) \nonumber \\
   =-{\rm cs}(a){\rm cs}(2a)[{\rm dn}(x+a)+{\rm dn}(x-a)] \nonumber \\
   +{\rm cs}^2(a){\rm dn}(x).
\end{eqnarray}

On the other hand, in the case of the MC model \cite{DKYF_PRE2006}, the
following identities are required for deriving the JEF
solutions \cite{kls}

\begin{eqnarray} \label{Identity4}
   m{\rm sn}^2(x)[{\rm sn}(x+a)+{\rm sn}(x-a)] \nonumber \\
   ={\rm ns}^2(a)[{\rm sn}(x+a)+{\rm sn}(x-a)] \nonumber \\
   -2{\rm cs}(a){\rm ds}(a){\rm sn}(x),
\end{eqnarray}

\begin{eqnarray} \label{Identity5}
   m{\rm cn}^2(x)[{\rm cn}(x+a)+{\rm cn}(x-a)] \nonumber \\
   =-{\rm ds}^2(a)[{\rm cn}(x+a)+{\rm cn}(x-a)] \nonumber \\
   +2{\rm cs}(a){\rm ns}(a){\rm cn}(x),
\end{eqnarray}

\begin{eqnarray} \label{Identity6}
   m{\rm dn}^2(x)[{\rm dn}(x+a)+{\rm dn}(x-a)] \nonumber \\
   =-{\rm cs}^2(a)[{\rm dn}(x+a)+{\rm dn}(x-a)] \nonumber \\
   +2{\rm ds}(a){\rm ns}(a){\rm dn}(x).
\end{eqnarray}

\section*{Acknowledgements}
PGK gratefully acknowledges support from NSF-DMS-0204585,
NSF-DMS-0505663 and NSF-CAREER.  This work was supported in
part by the U.S. Department of Energy.

\end{document}